\documentstyle[twocolumn,prd,aps]{revtex}  % DON'T CHANGE
\input epsf
\begin{document}                % INITIALIZE - DONT CHANGE
\title{Adaptive Event Horizon Tracking and Critical Phenomena in Binary
Black Hole Coalescence}
\author{Scott A. Caveny and Richard A. Matzner}
\address{Center for Relativity, The University of Texas at Austin, Austin,
TX 78712-1081 }
% \author{}   % Use this and the next line only if there is a second
% \address{Another University, etc.}  % address. (Remove the left % marks)
%
\maketitle
\begin{abstract}                % DON'T CHANGE THIS LINE
This work establishes critical phenomena in the topological
transition of black hole coalescence. We describe and validate
a computational front tracking event horizon solver, developed for generic
studies of the black hole coalescence problem. We then apply this to
the Kastor - Traschen axisymmetric analytic solution of the extremal
Maxwell - Einstein black hole merger with cosmological constant.  The
surprising result of this computational analysis is a power law scaling  of
the
minimal throat proportional to time. The minimal throat connecting the
two holes obeys
this power law during a short time immediately at the beginning
of merger. We also confirm the behavior analytically. Thus, at least
in one axisymmetric situation a critical phenomenon exists. We give
arguments for a broader universality class than the restricted
requirements of the Kastor - Traschen solution.

\end{abstract}
\section{Introduction}               % Introduction goes below.
There is a two-fold motivation for this work. On one hand, as
argued in \cite{eh1:}, \cite{eh2:}, \cite{eh3:}, and \cite{eh4:}, a robust
event
horizon solver provides an abundance of intuition and information
in numerical analysis of the binary black hole coalescence
problem. Alternatively, several open questions concerning the
dynamics of  black hole event horizons in the strong field domain
remain unanswered in the literature. In this work,
these motivations are pursued in development of a robust event horizon
solver
that is capable of establishing an intriguing  analogy between black holes
undergoing merger and fluid droplets undergoing bifurcation.

A theorem due to Penrose (cited in \cite{MTW:}) establishes that the event
horizon of a black hole is generated by null geodesics with no
future end point. Specifically,
\begin{enumerate}
\item Followed into the past, a null geodesic can only leave the event
horizon at a special event denoted a caustic.
\item Through each non-caustic event of the event horizon there is
a unique null geodesic.
\end{enumerate}
As demonstrated below, these properties of the generators yield a rich
structure for the dynamical evolution of the event horizon.

For example, it can be shown by way of the no hair theorem that
the topology of sections of the event horizon for a stationary black
holes is necessarily spherical. It was believed, at first, that
in the dynamical strong field regime similar behavior persists. In
a surprising result, Hughes et al., and others found numerically a
momentarily toroidal section of the event horizon in the
nonstationary case of the gravitational collapse of a ring of
particles \cite{eh1:}. Subsequent work by Shapiro, Teukolsky and Winicour
\cite{wini1:}
and others  \cite{jacobson:}, \cite{wald3:}, \cite{siino1:},
\cite{siino2:} has shown that this result is correct and is
currently conjectured to be generic for asymmetric gravitational
collapse \cite{wini2:}. Briefly, in an exactly future asymptotically
stationary spacetime the horizon generators can be shown to merge in
the asymptotic past at a zero dimensional point, which is unstable
to perturbation of this zero dimensional locus of generators.
Thus, horizons that
are dynamical in the asymptotic future yield higher dimensional loci
--- curves and surfaces --- of merged generators. Consequently,
there exists a higher genus horizon for some interval in the evolution.
Numerical
and exact studies have presented similar evidence in
\cite{wini2:} leading to the conjecture that event horizons of black holes
are unrestricted in their genus
\cite{wini2:}.

Related work on the phenomenology of black hole event horizons
concerns the differentiability of the two dimensional membrane of
the event horizon sections \cite{chrusciel1:}, \cite{chrusciel2:},
\cite{chrusciel3:}. Creases and caustics of the event horizon,
where the horizon is no longer differentiable, serve an important
role both in the topology of single black hole event horizons and
in the merger of multiple black holes into a single horizon. Where
the event horizon of a black hole undergoes a change in topology,
the surface of section becomes strictly singular \cite{wini2:}.
Thus, where two or more sections of the event horizons merge
to form a single horizon each surface of section is singular
and no longer differentiable at the points of merger
\cite{siino1:}, \cite{siino2:}. These special crease and caustic
points of the horizon are currently unrestricted in their presence
on the horizon. For example, it can be shown \cite{chrusciel1:}
that working only with the definition of a black hole event
horizon in terms of null geodesic generators having no future end
point, it is possible to construct event horizons that are nowhere
differentiable. Physically this can be approximated by a ``cloud of sand''
falling into a large black hole: Each ``grain'' falling into the hole
generates a caustic that terminates where the grain crosses the horizon.

Both of these considerations are problematic to numerical studies
--- particularly ones that seek a generic solution or work from
the finite difference approximation. These interesting
phenomenological aspects of the caustics and crease sets of black
hole event horizons undergoing merger or other topological
transitions are an open field of research.

Here we concentrate on
the development of a generic numerical black hole event horizon
tracking method.  It is possible to generically solve this
numerical problem, and the solution is presented in \cite{cavenyetal:}.
The computational technique uses the finite difference approximation,
thus assumes a degree of differentiability which will not be achieved at
caustics, creases, etc. Nevertheless, we find that the approach provides
very good approximation in those cases; e.g., gives surfaces very close (and
convergently close) to analytical expectations.
In particular, problems associated to differentiability of the
horizon that arise due to creases and caustics of the horizon can
be handled on a case by case basis. One such approach, advocated here,
is the  incorporation of adaptive
mesh refinement (AMR). Incorporating AMR into a numerical
method allows special points, such as crease sets or caustics, to
be resolved to within machine epsilon --- a resolution that is
typically sufficient to determine the global behavior of the
phenomena. But it remains that
special attention is required if very fine scales
are to be resolved.

A subject related both to the need for AMR in numerically
tracking black hole
event horizons and to the presence of crease or caustic
points in continuum black hole event horizons, is the question
of power law scaling of black hole event horizons undergoing
topological transitions. To motivate the possibility of this
effect, it is useful to  consider the `membrane paradigm' of
black hole event horizons.
In this approach the event horizon sections
are considered in analogy to a two dimensional fluid, such as the
surface of a liquid droplet and in fact, the dynamics of the
sections of the horizon have been  shown to obey
evolution equations analogous to those of a two dimensional
fluid \cite{membrane:}. Accordingly, the event horizon of a black hole can
be
viewed as a distinct dynamical physical entity within a spacetime.
Further, the membrane paradigm equations were demonstrated for
numerically detected black hole event horizons in \cite{eh5:} and,
as demonstrated in later sections, offer insight into the dynamics
governing the numerical evolution of the critical membrane.

The formal analogy between black hole event horizons and the
surface of a fluid droplet, spelled out by the membrane paradigm,
suggests that the critical phenomena associated to fluid droplet
bifurcation are also present in the binary black hole coalescence
problem. Study of the fluid problem typically establishes
power law scaling of the throat using both numerical and exact
analysis of the Navier Stokes equation governing the dynamics; in
almost all of those studies, use is made of AMR \cite{eggers:}. The
successes of
the membrane paradigm and fluid studies of droplet bifurcation
suggest that AMR applied to black hole event horizon solvers
will produce analogous black hole results.

In section II
we present a new AMR method for tracking black hole event horizons.
We call our method the comoving front tracking method and notate it
here as \texttt{cmft}. In section II the specifics of our method are
shown in detail by building on the work of  \cite{eh4:}, which
addressed the problem of numerically tracking black holes as a
computational front tracking problem. The accuracy of an implementation
of our method is shown in sections II. A. and B. which consider
the case of Kerr - Newman black holes with and without coordinate
deformations of the source. Section III applies our method to the
Kastor - Traschen solution \cite{kt:}, describing the merger of two charged
black holes in a spacetime with cosmological constant. In that section
we study the minimal radius of the neck connecting the two black holes
immediately following merger. We find, in analogy to the case
of fluid droplets undergoing bifurcation, that the minimal radius
of the neck undergoes power law scaling with the minimal throat
radius proportional to time. In section IV, we summarize and discuss
our conclusions.

\section{Comoving Front Tracking}
Adaptive mesh refinement is a numerical method developed
generically for hyperbolic systems \cite{berger:}. The method was
first applied in numerical relativity, with considerable success,
by Mathew Choptuik, in study of the problem of black hole
formation under the gravitational collapse of  massless scalar
fields \cite{choptuik:}. The method embodies Brandt's rule of
numerical analysis: Computational resources should be applied
proportionally to the physical processes involved \cite{brandt:}.
As the method's name implies, adaptive mesh refinement (AMR) is
typically applied where solution or field variables require
variable resolution over the computational mesh: Some sectors of
the grid require higher resolution (where the field variables
undergo interesting changes), while other sectors undergo less
activity and so need not be computed to high accuracy. There are
two distinct approaches to AMR. In one approach, there is one grid
of coarsest resolution that is fixed but within which higher
resolution grids are nested according to the behavior of the
solution. In the alternate method of AMR, which is the one used
here, there is one grid that moves with the solution. Such methods
have been applied with great success to other fields of
computational physics, particularly in problems involving the
motions of fronts, such as those found across phase transitions.

To understand the \texttt{cmft} method
for tracking black hole event horizons, it is useful to first
consider the method as applied in a fixed mesh
to the problem
of tracking black hole event horizons in \cite{eh4:}, although
those studies did not consider the application of adaptive meshes.
Since the event horizon is a null surface (away from caustics) one
must study the null geodesic equations; we solve the null geodesic
equations by solution of an eikonal equation.
In the front tracking approach a clever coordinate system, adapted
to the black hole's surface of section $\Gamma$, is chosen. Let
$\left\{\sigma_{i}\right\}$ be such a coordinate system. Then the
front $\Gamma$ of a single level set of $S$, where $S$ solves the
eikonal equation
\begin{equation}
g^{a b} \partial_{a}S \partial_{b}S = 0,
\end{equation}
can be tracked by elimination of one coordinate;
vis,
\begin{equation}\label{coords}
S\left(x^{i},t\right) = \sigma_{1} -
u\left(\sigma_{1},\sigma_{2},t\right) = 0.
\end{equation}
In terms of these coordinates, the eikonal equation written in the
ADM variables
\begin{equation}\label{eik}
\partial_{t}S = \beta^{i}\partial_{i}S \pm
\sqrt{\partial_{j}S\gamma^{jk}\partial_{k}S}
\end{equation}
becomes
\begin{equation}\label{surf}
\partial_{t} u = -\beta^{1} + \beta^{I}\partial_{I}u \pm
\alpha \sqrt{\gamma^{11} + 2\gamma^{1 I}\partial_{I}u +
\partial_{I}u \gamma^{IJ}\partial_{J}u}.
\end{equation}
Here $I=2,3$ and $\gamma^{IJ}$ is the two dimensional metric
obtained by the choice of coordinates.

Several comments are relevant. First, this method
considerably extends the non - linearity of the eikonal equation:
The geometric variables $\alpha, \beta^{I}, \gamma^{IJ}$ are each
functions of the coordinates $x^{i}$ and consequently, the
geometric variables are also functions of the grid function $u$;
e.g., $ x^{i} = x^{i}\left(\sigma^{k}\right)\equiv
x^{i}\left(u,\sigma^{I}\right)$. That is, whereas the eikonal
(\ref{eik}) was a hyperbolic equation of the general nonlinear
form
\begin{equation}
\partial_{t} S = F\left(t,x,\partial_{x}S\right),
\end{equation}
equation (\ref{surf}) is a hyperbolic equation of the general form
\begin{equation}
\partial_{t} u = G\left(t,x,u,\partial_{x}u\right).
\end{equation}
As a direct consequence, while the nonlinearity of the
gradient of $u$ is given in terms of the root in (\ref{surf}), the
nonlinearity in terms of $u$ typically cannot be classified;
particularly when the geometric variables are only provided
numerically.  For example, in numerically generated spacetimes, a
case for almost all problems of interest, the geometric variables
$\alpha, \beta, \gamma^{IJ}$ are defined on a global  grid
$\mathcal{G}$ of $N^{3}$ points
\begin{equation}
x^{i}\left(k^{i}\right) = s^{i} k^{i} + x^{i}\left(0\right),
\end{equation}
where the integer $k^{i}$ satisfies $1 \leq k^{i} \leq N$ for each
$i$. The front tracking method then generically requires
interpolation methods since the two - dimensional mesh $G^{0}$ used
for the finite difference approximation of the two - dimensional
surface of sections of the horizon $\Gamma$ will not in general
coincide with the
points of the global grid $\mathcal{G}$. Update of the grid
function $u$ then requires knowledge of the geometric variables
$\alpha, \beta, \gamma^{IJ}$ at points that do not lie on
$\mathcal{G}$, which can only be found by interpolation. A
further complication of the front tracking method relates to the
question of the coordinates $\left\{\sigma_{i}\right\}$. As
described in \cite{eh4:} the method requires different coordinate
systems depending on the physical processes involved. [In fact,
the original studies \cite{eh4:} found dependencies and sensitivities of
results on the choice of coordinates, which is an unappealing
feature of the method, particularly from the viewpoint of general
relativity.] For example, in the case of a single hole with static
topology a spherical coordinate system is typically sufficient.
However, for the case of head on binary black hole merger a
cylindrical coordinate system was employed \cite{eh4:}, while it
is unclear what global coordinates should be chosen, or if there
is even one set of coordinates which can be used for the problem
of two black holes undergoing inspiral to merger.

The method of adaptive mesh front tracking developed here
addresses several of the complications of the method described
above. First, within our approach we fix the choice of
coordinates as spherical coordinates although the \texttt{cmft}
method is not contingent on this choice. Instead,
spherical coordinates are chosen due to their
boundary conditions, which are advantageous
in implementation. Note that while
spherical coordinates clearly cannot handle evolution into any topologies
beyond the genus zero, $S^2$ topology of stationary black hole
event horizons, the \texttt{cmft} method compensates for
this by allowing the mesh to adaptively track
black hole event horizons that are stationary in their asymptotic
past and future; more importantly, through refinement, to
naturally detect the onset of topology change, where (strictly) the
surface becomes singular. In such circumstances, the surface can
be monitored to within arbitrary proximity of the transition; and then
continued past the transition by applying the code individually to
the resulting black holes.

To make the method precise, at some fixed time level $t = t_{n}$
let $c^{i} = \left<x\right>^{i}$ be the average value of $M^{2}$
points distributed over a surface $\Gamma$ having $S^{2}$ topology
and satisfying a level set condition $S\left(\Gamma\right) = 0$.
That is, in a two dimensional mesh $G^{n}$ of $M^2$ points $x^{i}
= x^{i}\left(I,J\right) $ on $\Gamma$  where the integers
$I,J$ are $1 \leq I,J \leq M$ let
\begin{equation}
c^{i} = \left<x^{i}\right> \equiv
\frac{1}{M\left(M-1\right)}\sum_{I,J=1}^{I,J=M,M-1}x^{i}\left(I,J\right).
\end{equation}
Local coordinates on the surface $\Gamma$ can then be
written
\begin{equation}
x  -  c^{1} = r \cos\left(\phi\right)\sin\left(\theta\right)
\end{equation}
\begin{equation}
y - c^{2}= r \sin\left(\phi\right)\sin\left(\theta\right)
\end{equation}
\begin{equation}
z - c^{3} = r \cos\left(\theta\right).
\end{equation}
According to this choice $r = u\left(\theta,\phi,t_{n}\right)$ can
be updated in the two dimensional mesh $G^{n}$ according to a
finite difference representation of  equation (\ref{coords})
expressed in the spherical coordinates
That is, the form of (\ref{coords}) is chosen here as
\begin{equation}
r = u\left(\theta, \phi,t\right) = \sqrt{\left(x -
c^{1}\right)^{2} + \left(y - c^{2}\right)^{2} + \left(z -
c^{3}\right)^{2}}.
\end{equation}

The coordinates $x,y,z$ are defined both in terms of the global
grid $\mathcal{G}$ and the $M^{2}$ mesh $G^{n}$ having center
$c^{i}$. By comparison, the grid function $u$ is defined locally
on the comoving mesh $G^{n}$. In the case that the geometric
variables $\alpha, \beta^{i}, \gamma^{ij}$ are provided
numerically from solution on the global $N^{3}$ grid $\mathcal{G}$,
interpolation of those variables onto the local $M^{2}$ surface
$\Gamma$ is required for update of $u$. This is a generic feature of the
front tracking method. Further, in
our method in one cycle of the construction, the
radial coordinate $u$ will evolve in accordance with the update since
the surface center $c^{i}\left(t\right)$ will also
update: $c^{i}\left(t\right) \rightarrow c^{i}\left(t+dt\right)$.
This update of the surface center corresponds to creation of a new
$M^{2}$ mesh $G^{n} \rightarrow G^{n+1}$ and therefore we need
a numerical change of variables $u^{n+1}_{IJ} \rightarrow
u^{n+1}_{I'J'}$ between the two meshes. This change of variables
will always, irrespective of the nature of the geometrical
variables, require interpolation of the grid function $u$ from the
one grid, $G^{n}$, to the other, $G^{n+1}$, since the points of
those grids do not coincide in general. This procedure can become
in practice a very detailed and intricate computational step since
it amounts to passing data from one spherical mesh to another
spherical mesh and the procedure can potentially fall prey to grid
tangling effects. Most of the intricacy of the data passing $G^{n}
\rightarrow G^{n+1}$ is restricted to the choice of spherical
coordinates. For the purposes of this work ordinary second order
interpolation proves sufficient both for grid passing and for
interpolation of the geometric variables onto the surface.
Finally, the iterated Crank Nicholson scheme is generically well
suited as a finite difference approximation for the partial
differential equation (\ref{surf}). Studies were conducted with
other schemes, such as the method of lines used by other
researchers in the front tracking problem, but superior
performance was found with the iterated Crank Nicholson
method. Discussion of this finite difference approximation can be
found in \cite{icn:}.

In summary, a pseudo code expression for one complete update of
the comoving geometry is then:
\begin{center}
\texttt{Pseudo-Code:} \texttt{A Complete Update Iteration }
\begin{itemize}
\item{} \texttt{Load $\left\{x^{n}_{IJ}\right\}$}
\item{} \texttt{Build $c^{n}$ and  $u^{n}_{IJ}$}
\item{} \texttt{Interpolate $\gamma$ onto $\Gamma$}
\item{} \texttt{Update $u^{n}_{IJ} \rightarrow u^{n+1}_{IJ}$}
\item{} \texttt{Update $c^{n} \rightarrow c^{n+1}$}
\item{} \texttt{Pass data  $u^{n+1}_{IJ} \rightarrow u^{n+1}_{I'J'}$}
\item{} \texttt{Return $\left\{x^{n+1}_{I'J'}\right\}$}
\end{itemize}
\end{center}

\subsection{Kerr-Newmann Black Holes}
\begin{figure}% Imported eps example.
\epsfxsize=8cm
\centerline{\epsfbox{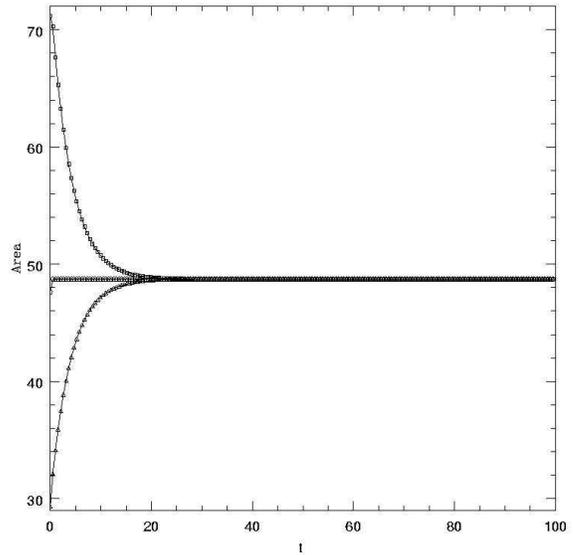}}
\vspace{0.5cm}
\caption{\texttt{CMFT} detection and tracking, in units of $M$, of
a nonspinning $M = 1$ black hole. Here area is given in units of $M^2$,
and increasing $t$ (units of $M$) corresponds to
propagation into the past.
}
\label{f:0}
\end{figure}

In this section the accuracy of an implementation of the
\texttt{cmft} method is considered in detail. For more detailed
discussion of
the signatures
of black hole event horizons in the eikonal equation see
\cite{cavenyetal:} and \cite{eh4:}. Briefly, since the event
horizon of a black hole
is a global structure of spacetime, its detection cannot be
determined without the complete history of the Cauchy
evolution of the spacetime. However, as shown below,
the horizon is a critical
outgoing null
surface that neither expands to infinity nor collapses into the
gravitational
singularity. Numerical event horizon solvers have typically employed this
property as the signature behavior of event horizon detection
and tracking. In these approaches, the space of outgoing null surfaces
propagated into the past is surveyed  for
evidence of the critical behavior of the horizon. Since
outgoing null data followed into
the past tend to approach the horizon to great precision these
methods typically
produce highly accurate approximations of the event horizon itself.
For example, in figure (\ref{f:0}) we show outgoing data propagated
into the past in a background spacetime contatining a spherically
symmetric black hole. In that figure three classes of data are considered:
Null data initially interior to the black hole event horizon,
null data initially exterior to the
black hole event horizon and null data that is exactly on the black hole
event horizon. In all cases, we see
that outgoing null data propagated into the past approaches the black hole
event horizon to high accuracy.

\begin{figure}% Imported eps example.
\epsfxsize=8cm
\centerline{\epsfbox{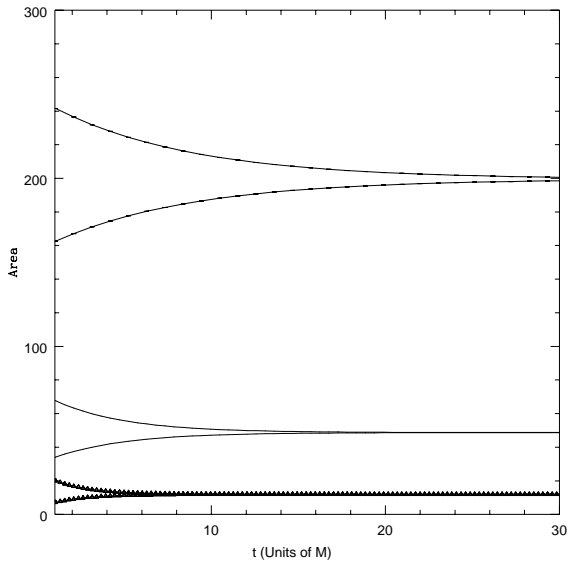}}
\vspace{0.5cm}
\caption{\texttt{CMFT} detection and tracking, in units of $M$, of
nonspinning black holes: $M = 1/2, 1, 2$.
Here increasing $t$ corresponds to
propagation into the past.
} \label{f:1}
\end{figure}

To establish the accuracy of our implementation, we consider first
the case of stationary, spinning black holes, which is
completely described by the Kerr - Newmann axisymmetric
solutions of Einstein's equation. According to Carter's
theorem \cite{carter:}, this family of solutions is the unique,
asymptotically flat, stationary and axisymmetric black hole
solutions of the vacuum equations. As such, it is sufficient to
consider this class of solutions in an account of stationary black
hole event horizons. It is convenient to the discussion to make
use of the Kerr-Schild form for the metric \cite{deirdre:}:
\begin{equation}
g^{ab} = \eta^{ab} - 2 H l^{a}l^{b}.
\end{equation}
Here $\eta_{ab} = \mathrm{diag}\left(-1,1,1,1\right)$ is
Minkowski's metric, $H$ is a space time scalar, and $l^{a}$ is an
ingoing null vector with respect to both the Minkowski and full
metric. The Kerr solution is the two parameter family of solutions
such that
\begin{equation}
H = \frac{M r^{3}}{r^{2} + a^{2} z^{2}}
\end{equation}
and
\begin{equation}
l^{t} = -1,
\end{equation}
\begin{equation}
l^{x} = \frac{r x + a y}{r^{2} + a^{2}},
\end{equation}
\begin{equation}
l^{y} = \frac{r y - a x}{r^{2} + a^{2}},
\end{equation}
\begin{equation}
l^{z} = \frac{z}{r},
\end{equation}
\begin{equation}
r^2 = \frac{1}{2}\left(\rho^{2} - a^{2}\right) +
\sqrt{\frac{1}{4}\left(\rho^{2} - a^{2}\right) + a^{2} z^{2}}
\end{equation}
where
\begin{equation}
\rho^2 = x^{2} + y^{2} + z^{2}.
\end{equation}

\begin{figure} % Imported eps example.
\epsfxsize=8cm
\centerline{\epsfbox{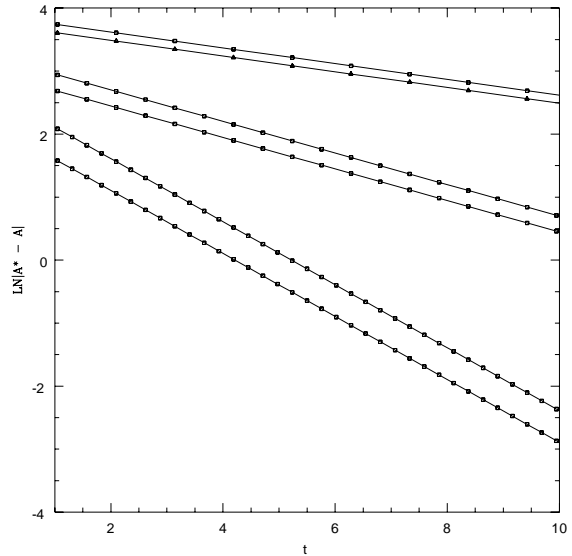}}
\vspace{0.5cm}
\caption{\texttt{CMFT} analysis of $e$-folding time, in units of
$M$, for nonspinning black holes of different masses.
This figure shows pairs of interior
and exterior data converging to the horizon, where the pairs are top
to bottom $4M$, $2M$, and $M$.
Here increasing $t$ corresponds to
propagation into the past.} \label{f:2}
\end{figure}

Two points are worth mentioning. Firstly, the parameter $M$
corresponds to the gravitational mass of the source, while the
parameter $a$ corresponds to the source's spin: It can be shown
that observers at asymptotic infinity measure the angular
momentum of the source to be $J = a M $. Secondly, the cartesian
coordinates are chosen such that the $z$ axis is
aligned with the direction of spin and so is an axis of symmetry.

\begin{figure} % Imported eps example.
\epsfxsize=8cm
\centerline{\epsfbox{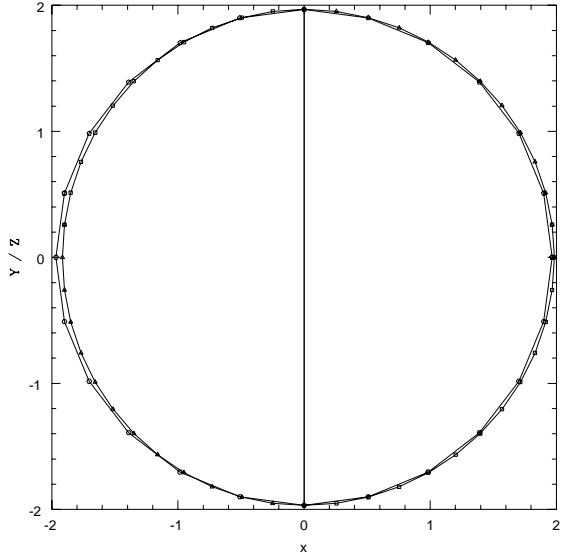}}
\vspace{0.5cm}
\caption{\texttt{CMFT} tracking of spinning Black hole $a/M =
0.25$.} \label{f:3}
\end{figure}

The Kerr family of solutions is interesting from the perspective
of tracking black hole event horizons since the solutions actually
contain two event horizons; with one nested interior to the other.
Furthermore, the solutions have a ring curvature singularity
located at $r = z = 0$. In the \texttt{cmft} method, provided the
event horizon is approached from the exterior, neither of these
features requires special attention.

\begin{figure}% Imported eps example.
\epsfxsize=8cm
\centerline{\epsfbox{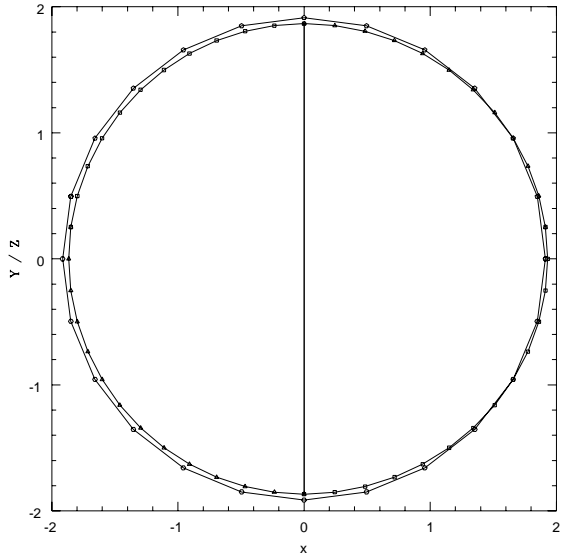}}
\vspace{0.5cm}
\caption{\texttt{CMFT} tracking of spinning Black hole $a/M =
0.5$} \label{f:4}
\end{figure}

It is analytically known that the outermost horizon of a spinning black
hole is located at the surface
\begin{equation}
r_{+} = M + \sqrt{M^{2} - a^{2}}
\end{equation}
with an area
\begin{equation}
A = 4 \pi \left({r_{+}}^{2} + a^{2}\right).
\end{equation}
We will now validate our \texttt{cmft} codes' ability to determine
these values.

\begin{figure}  % Imported eps example.
\epsfxsize=8cm
\centerline{\epsfbox{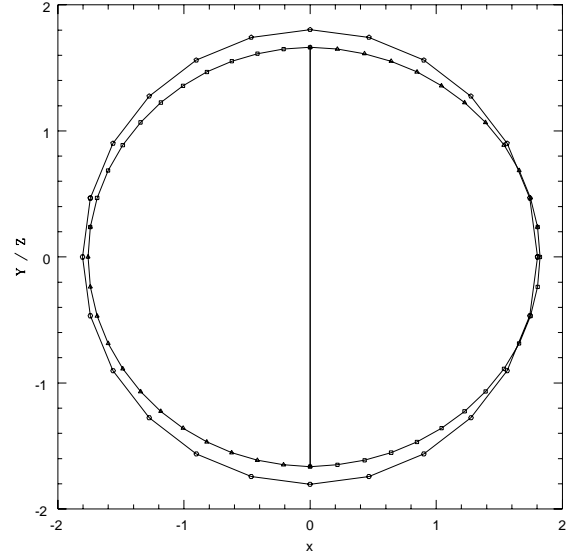}}
\vspace{0.5cm}
\caption{\texttt{CMFT} tracking of spinning Black hole $a/M =
0.75$} \label{f:5}
\end{figure}

We start with nonspinning black
holes. With $a = 0 $, figure (\ref{f:0}) shows the signature of
the black hole event horizon for outgoing null data propagated
backwards in time. Figure (\ref{f:1}) similarly shows this effect for
outgoing data and includes  variation in the mass with the cases
of $M = 1/2, 1, 2$ considered.
\begin{figure}  % Imported eps example.
\epsfxsize=8cm
\centerline{\epsfbox{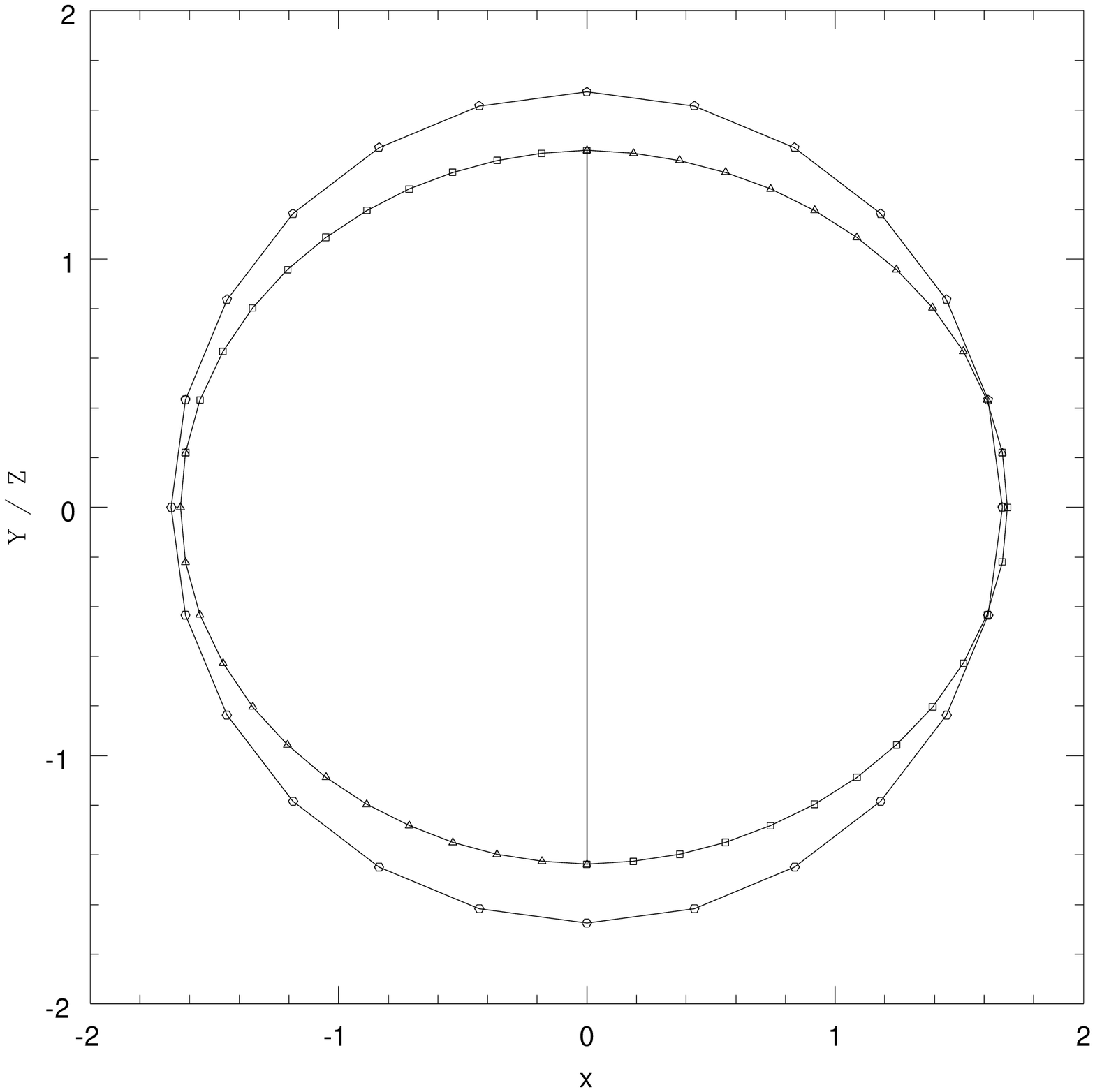}}
\vspace{0.5cm}
\caption{\texttt{CMFT} tracking of spinning Black hole $a/M =
0.9$} \label{f:6}
\end{figure}
In all cases interior and exterior data converge exponentially to
the event horizon as predicted by the exact solution. The $e$ -
folding time of this exponential behavior (which can be shown in
perturbation theory to satisfy $\gamma = \left(4M\right)^{-1}$ for ingoing
Eddington - Finkelstein coordinates) is shown numerically in
figure (\ref{f:2}). The slope of each line is $\left(4 M\right)^{-1}M$ in
agreement with the perturbation prediction.

\begin{figure}  % Imported eps example.
\epsfxsize=8cm
\centerline{\epsfbox{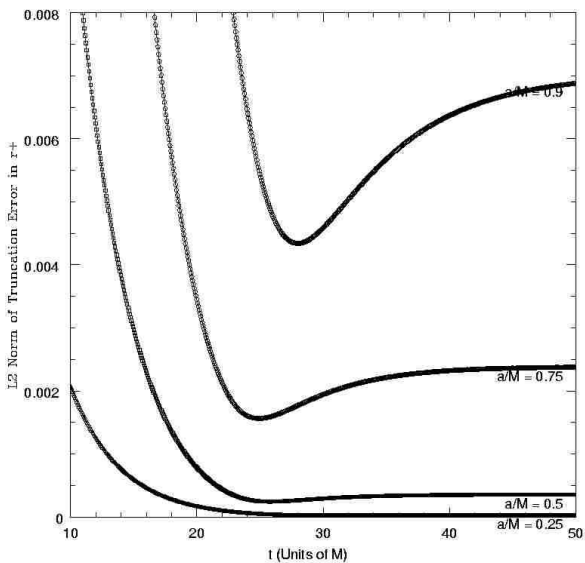}}
\vspace{0.5cm}
\caption{Scaling of truncation error, in units of $M$,  of $r_{+}$ for a
spinning Black hole $a/M = 0.25$, $0.5$, $0.75$, $0.9$ using \texttt{CMFT}
tracking. Here increasing $t$ corresponds to
propagation into the past.} \label{f:7}
\end{figure}

We turn now to consideration of spinning black holes.
The ellipsoidal geometry of sections of spinning black hole event
horizons are demonstrated in figures (\ref{f:3}), (\ref{f:4}),
(\ref{f:5}), (\ref{f:6}), which show the intersection of the
event horizon section both with a $\phi = \pi$ slice and with a $ \theta
= \pi / 2$ slice. These figures are the asymptotic limit cross
sections of spheres tracked backwards until the area achieves
stationarity. They are thus approximations of the black hole event
horizons.
The figures are drawn for $a/M = 0.25, 0.5, 0.75,$ and
$0.9$.  As can be seen there is a slight asymmetry (an error)
of the resulting surfaces. The accuracy of these results is shown in figure
(\ref{f:7}), which contains the L2 norm of the numerical
truncation error $e_{r}$ of
the $r_{+}$ function calculated over the surface of the final
state. With $\hat{r}_{+}$ denoting the finite difference approximation
of the continuum funtion $r_{+}$, the truncation error is defined
to be
\begin{equation}
e_{r} = \hat{r}_{+} - r_{+}.
\end{equation}
In figure(\ref{f:7}) time in units of $M$ is measured
increasing into the past.

\begin{figure}% Imported eps example.
\epsfxsize=8cm
\centerline{\epsfbox{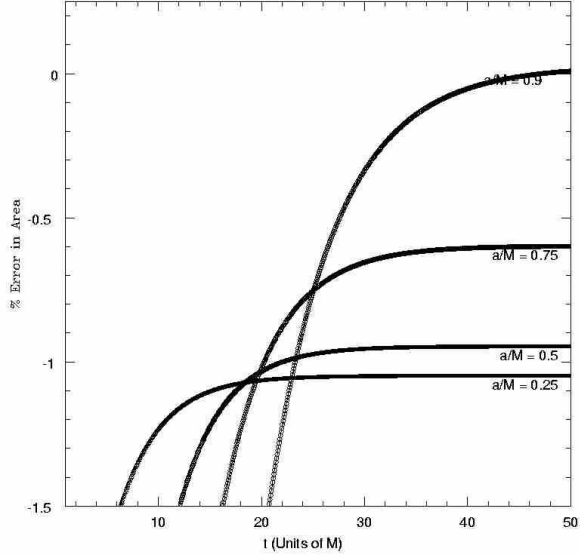}}
\vspace{0.5cm}
\caption{ Percent error in area for spinning black hole $ 0 < a/M < 0.9$
 using \texttt{CMFT} tracking. Here increasing $t$ corresponds to
propagation into the past.} \label{f:9}
\end{figure}

\begin{figure}% Imported eps example.
\epsfxsize=8cm
\centerline{\epsfbox{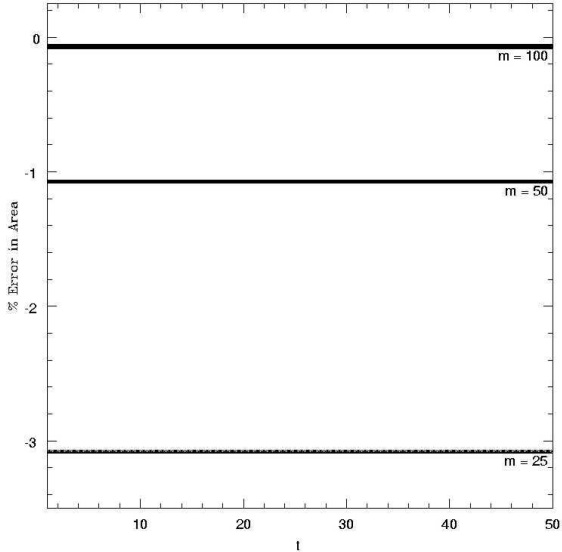}}
\vspace{0.5cm}
\caption{Scaling
of truncation error in area for spin zero black hole with surface
resolutions of $m^{2}$, with $m = 25$, $50$, $100$. Here increasing $t$
corresponds to
propagation into the past.} \label{f:10}
\end{figure}

Figure (\ref{f:9}) shows several values of the percent error in
the area for a fixed unit mass black hole with spin parameter $ 0
\leq a \leq 0.9$. These figures were generated using a
numerical resolution of $m^{2}$ with $m = 50$ on the detected
event horizon surface section\footnote{Throughout the studies in
this work an area calculation algorithm first proposed in
\cite{baum:} is used. In this method the metric $h$ of the space
time is determined on the surface $\Gamma$ and the area is
calculated using numerical integration of  $A = \int d\theta d\phi
\sqrt{h}$. Also, note carefully that lower case {\it} is a number, giving
the resolution of the surface discretization. {\it M} is a mass.}.
The truncation error of the numerical integration of
the area scales at least as well as ${\mathcal{O}}\left(h^2\right)$
as required for second order convergence. This scaling of the
truncation error is shown in figure (\ref{f:10}), which shows the
percentage error of the area of a nonspinning black hole for
resolutions of $m^2$ with $m = 25, 50 , 100$.

In terms of the consistency of these results for the truncation
error of $r_{+}$ and the percent errors of the calculated areas,
note that there are at least two sources of systematic error in
the numerical calculation of the area of any surface. One source
of error is the accuracy with which each point of the surface is
known, while the second source of error is the finite resolution
of the discrete version of the surface. Let $A$ denote the
continuum area and $\hat{A}$ be the discrete version of the
surface calculated using a resolution of $m^{2}$. Then by the
finite difference approximation
\begin{equation}
A = \hat{A} + {\mathcal{O}}\left(1/m^{2}\right)
\end{equation}
for second order convergence. However, if the points of the
surface are systematically inaccurate $\hat{A}$ will not converge
to $A$ in the continuum limit; instead $\hat{A}$ will converge to
the bias of the implementation. This result suggests that the bias in
$\hat{A}$ is on the order of $0.1$ percent.

\begin{figure} % Imported eps example.
\epsfxsize=8cm
\centerline{\epsfbox{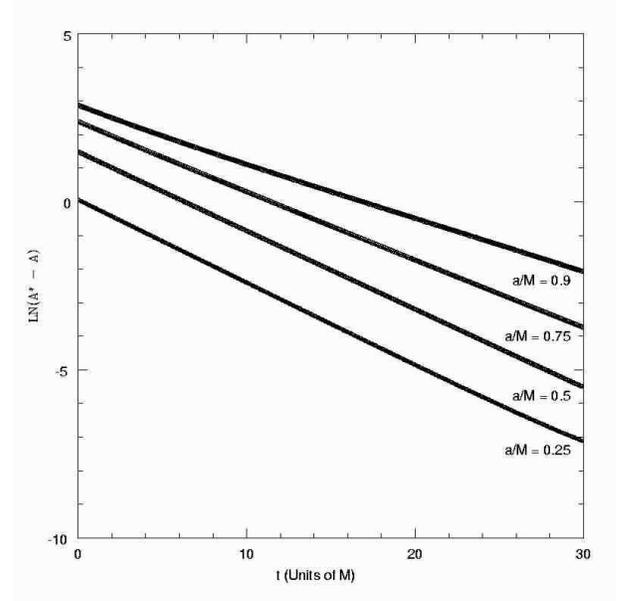}}
\vspace{0.5cm}
\caption{\texttt{CMFT} analysis of $e$-folding time, in units of
$M$,  for spinning black hole $0 < a/M < 0.9$. Here increasing $t$
corresponds to
propagation into the past.} \label{f:11}
\end{figure}

The $e$-folding time for relaxation of outgoing null data onto the
event horizon, or equivalently for formation of the solution
singularity in the eikonal, is shown in figure (\ref{f:11}) with
various values of the spin $a/M = 0.25,$ $0.5,$ $0.75,$ $0.9$.
Note that only for the cases of rapidly spinning black holes $a/M
\approx 0.9$ does the $e$-folding time differ appreciably from the
spin zero case of $\gamma = 1/ 4 M$. These results suggest using
the `rule of thumb' relaxation time of $\gamma \approx 1/ 4 M$ for
any spin with $a/M \leq 0.9$.


\subsection{Non-linear Coordinate Deformations}
The method of comoving adaptive meshes makes use of an intricate
numerical procedure: The passing and interpolation of data from
one spherical grid to another. To ensure that the computational
implementation of the method is
both accurate and robust, consider the case of the following
coordinate transformation:
\begin{equation}
t \rightarrow t' = t,
\end{equation}
\begin{equation}
x \rightarrow x' = x + b \cos \omega t,
\end{equation}
\begin{equation}
y \rightarrow y' = y + b \sin \omega t,
\end{equation}
\begin{equation}
z \rightarrow  z' = z
\end{equation}

A nonspinning black hole located at $x = y = z= 0$, is then seen
to rotate with angular frequency $\omega$ at a
radius of $r = b $ about the point  $x' = y' = z' = 0$. The
physics of this coordinate transformation is trivial and the area
of the source event horizon remains $A = 16  \pi M^{2}$. However,
from the perspective of the numerical implementation written in
the coordinates $\left(t',x',y',z'\right)$, the black hole appears
highly dynamical and moving with an angular velocity. As such, the
coordinate transformation  $\left(t,x,y,z\right) \rightarrow
\left(t',x',y',z'\right)$ applied to the Kerr - Newmann family of
analytic sources is a stringent test of an implementation of the
method of comoving adaptive meshes.

\begin{figure}% Imported eps example.
\epsfxsize=8cm
\centerline{\epsfbox{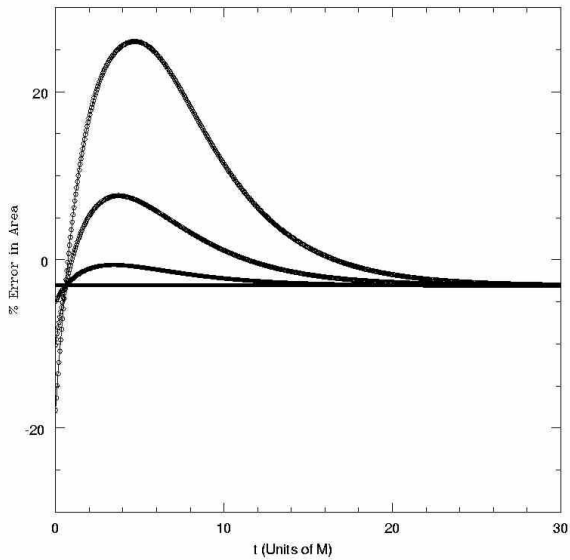}}
\vspace{0.5cm}
\caption{Percentage error in area of nonspinning black hole in
shifted coordinates: $b = 0, 1/2, 1, 3/2$. Here increasing $t$ corresponds
to
propagation into the past.} \label{f:20}
\end{figure}

\begin{figure} % Imported eps example.
\epsfxsize=8cm
\centerline{\epsfbox{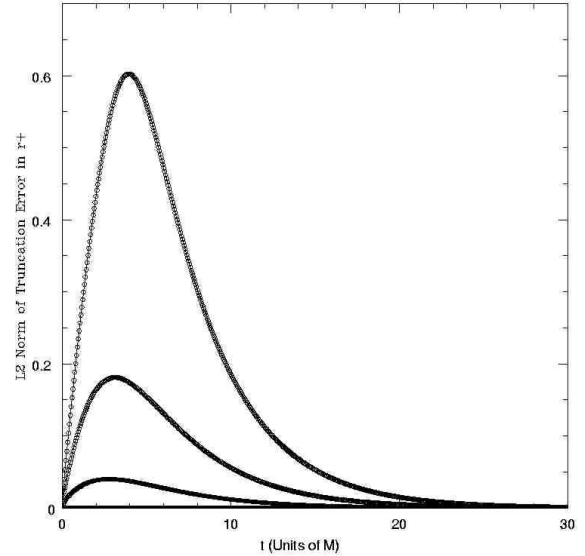}}
\vspace{0.5cm}
\caption{L2 norm
of truncation error, in units of $M$, of  $r_{+}$ for nonspinning
black hole in shifted coordinates: $b = 0$, $1/2$, $1$, $3/2$.Here
increasing $t$ corresponds to
propagation into the past.}
\label{f:21}
\end{figure}

Consider first the case of a linear shift of coordinates: $\omega
= 0, b \neq 0$. Figure (\ref{f:20}) shows the percent error
in the area of the detected horizon  versus time increasing under
propagation
into the past, while figure
(\ref{f:21}) shows the L2 norm of the error in the function
$r_{+}$ for $b = 3/2, 1, 1/2, 0$. Both of these figures are
generated using surfaces of $m^2$ points with $m = 25$ and
spherical initial data of radius $r = 2$ centered at the origin.
The source is that of a nonspinning unit mass black hole. The solution
ultimately settles down to the correct shift zero result, which
suggests that this implementation of the method is stable with
respect to translation of the source.

\begin{figure} % Imported eps example.
\epsfxsize=8cm
\centerline{\epsfbox{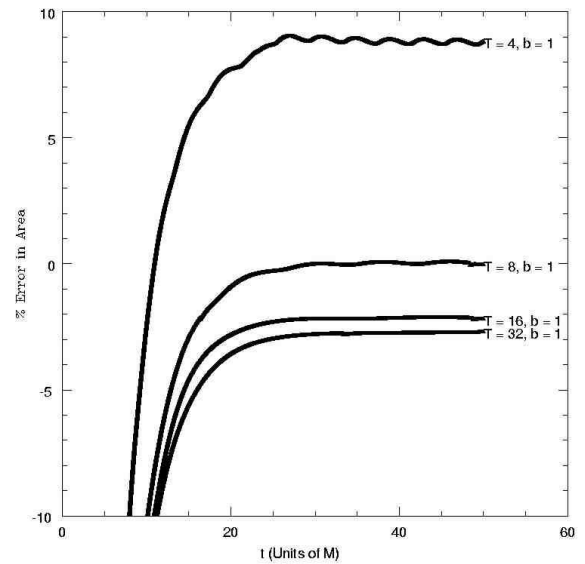}}
\vspace{0.5cm}
\caption{Percentage error  in area of  $a/M = 1/2$  black hole in
`wobbling' coordinates : $b = 1$, $\omega = 2 \pi / T=$ $\pi/ 2$,
$\pi / 4$, $ \pi /8$, $\pi / 16 $. Here increasing $t$ corresponds to
propagation into the past.} \label{f:22}
\end{figure}

\begin{figure} % Imported eps example.
\epsfxsize=8cm
\centerline{\epsfbox{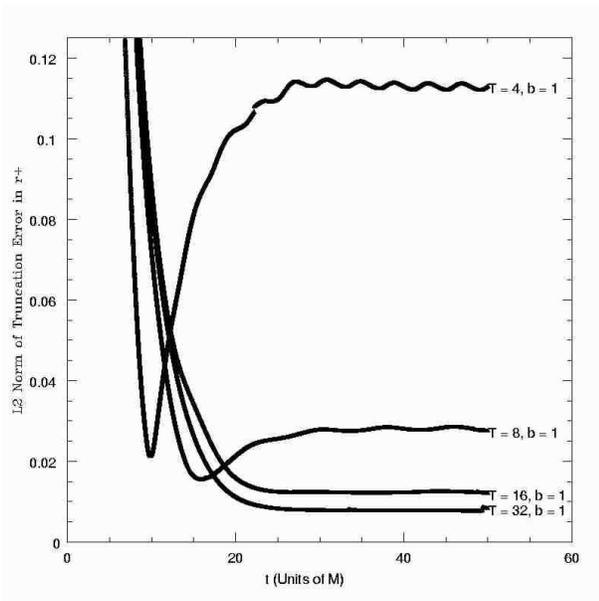}}
\vspace{0.5cm}
\caption{L2
norm of truncation error, in units of $M$,  of $r_{+}$ for $a/M =
1/2$ black hole in `wobbling' coordinates : $b = 1$, $\omega = 2
\pi / T $ $ = \pi/ 2$, $\pi / 4$, $\pi / 8$, $\pi / 16 $. Here increasing
$t$ corresponds to
propagation into the past.}
\label{f:23}
\end{figure}

Additional complexity is obtained by fixing $b = 1$ and varying
$\omega = 2 \pi / T$. Figure (\ref{f:22}) shows
the percentage error in the area of the detected horizon section,
while figure  (\ref{f:23}) shows the L2 norm of the error in
$r_{+}$. In these figures the surface used was that of $m^2$
points with $m = 25$, and spherical initial data of radius $r = 3$
located at the origin. Also here, the source considered in both
figures is a unit mass black hole with spin parameter $a = m/2$.
According to these results, this implementation of the method of
comoving adaptive meshes converges to the bias of the surface
resolution as $\omega \rightarrow 0$. Further, for time scales on
the order of the relaxation time $4M$, the error of the
implementation is fairly significant although it remains below
$10\%$. These result suggests that the particular implementation
is sensitive to coupling of the time scale of the event horizon's
dynamics. Further accuracy can be obtained by considering a more
stringent implementation.

\section{Symmetric Binary Black Hole Coalescence}
The  \texttt{cmft} method cannot
continuously monitor a topological transition in the event horizon
of a black hole. However, with high resolution or refinement the
method can detect the onset of a topological transition and come
arbitrarily close to the transition itself.

In this context we now consider the Kastor - Traschen
analytic solution of the Einstein - Maxwell $Q = M$ equations with
cosmological constant $\Lambda = 3 H^{2}$ \cite{kt:} \cite{kt2:}.
The solution is simply
\begin{equation}\label{kt}
ds^{2} = - U^{-2} dt^{2}+ U^{2}\left(dx^{2} + dy^{2} +
dz^{2}\right)
\end{equation}
where
\begin{equation}
U = - H t + \frac{M_{1}}{r_{1}} + \frac{M_{2}}{r_{2}}.
\end{equation}

$H=\pm\sqrt{\Lambda/3}$ and we set the coordinates and choose the sign
so that $H=1$. These are the choices made in ref \cite{Siino}.
Both refs. \cite{kt2:} and  \cite{Siino} show that with these choices the
black hole merger occurs in the future ({\it i.e.} as $t \equiv
t\uparrow$
increases into the future). The sign convention differs between
refs.  \cite{kt2:} and  \cite{Siino}; we follow  \cite{Siino}.

Here also, $M_{1}$ and $M_{2}$ denote the masses of the two holes
located at
$a_{1}$ and $a_{2}$ with
\begin{equation}
r_{1,2} = \sqrt{x^2 + y^2 + (z - a_{1,2})^{2}}.
\end{equation}

\begin{figure}% Imported eps example.
\epsfxsize=8cm
\centerline{\epsfbox{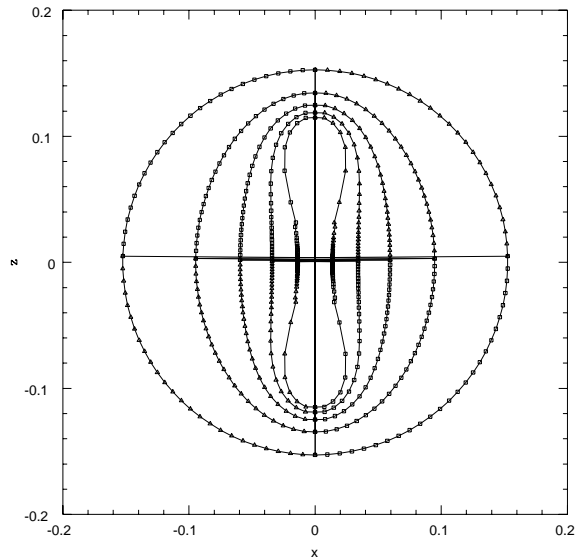}}
\vspace{0.5cm}
\caption{Onset of
topological transition in the Kastor Traschen solution using
\texttt{CMFT} tracking as the event horizon is tracked backwards
in time from the late time (outermost) horizon.} \label{g:1}
\end{figure}

We consider the equal mass case ($M_1=M_2=0.1$), symmetrically oriented
($a_{1} = -a_{2}  = a =0.1$) on the $z$-axis. These are the parameters
chosen in a calculation by Siino\cite{Siino}. The choice of relatively
small $M$ keeps the final (postmerger) single black hole horizon inside
the cosmological (deSitter) horizon. Ref. \cite{kt2:} demonstrates that
for small negative $t$ there is one component of the event horizon that
encompasses the origin; at somewhat earlier times there are two
separated components of the event horizon. Figure \ref{g:1} shows the
\texttt{cmft} front tracker result (running backward from a late time t
but still with $ t<0$ ), with an ``initial" guess equal to the late time
single apparent horizon. Refs \cite{kt2:} and \cite{Siino} estimate the
late time single horizon at the apparent horizon:
\begin{equation}
r_{bh} = -R_{bh}/(Ht) = -(1-2MH-\alpha)/(2 t H^2),
\end{equation}
where
\begin{equation}
\alpha = \sqrt{1-4MH},
\end{equation}
and this $M=M_1+M_2$. ($R_{bh}$ is a constant, though $r_{bh}$ is not.)

The coalescence occurs at a small negative
time estimated in \cite{kt2:} as: $t^{*} \approx  -R_{bh}/(2aH)<0$,
which for our parameters is $t^* = -0.763932$.  Following coalescence
the merged
black hole settles down to the final state of a charged black hole in a
spacetime with cosmological constant. Figure (\ref{g:1}) shows several
frames of a movie indicating bifurcation of the black hole event
horizon when tracked backwards in time from the outermost, guess
surface at the latest time. Note that the solution will eventually
break down when the grid function $u$ becomes multiple valued. Note
also that near the bifurcation the solution exhibits multiple time
scales. For example, away from the throat the solution appears
quasi-stationary, while at the throat the solution is clearly dynamic.

As we investigate the behavior of the horizon found by \texttt{cmft}
near
breakdown, we can look for evidence of power law scaling of the
topological
transition. In analogy to the topological transitions found in the
bifurcation (``pinch off") of fluid droplets, such power law scaling
would be
expected in the neck of the event horizon at the point and time of
merger, or
``pinch on" \cite{eggers:} \cite{goldstein:} (the opposite of ``pinch
off").
Using cylindrical
coordinates, let $\rho$ denote the radius of the throat connecting
the two black holes. By the axial symmetry of the solution, $\rho
= \rho\left(z,\phi, t\right) = \rho\left(z, t\right)$. At the
``pinch on" time, $t = t^{*}$, $\rho\left(0, t^{*}\right) = 0$. If
the solution exhibits power law scaling then
\begin{equation}
\rho = C \left(t - t^{*}\right)^{\gamma}
\end{equation}
for times after the ``pinch on".
\begin{figure}% Imported eps example.
\epsfxsize=8cm
\centerline{\epsfbox{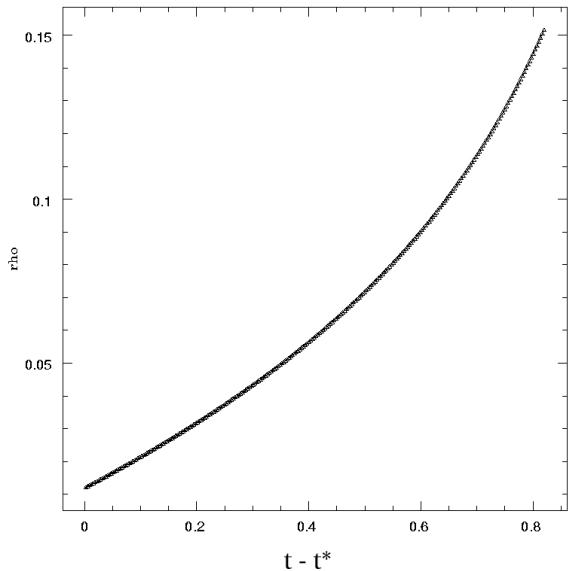}}
\vspace{0.5cm}
\caption{ Throat
radius $\rho$ versus $t- t^{*}$ just past the ``pinch on" at $t= t^{*}$
in the Kastor-Traschen solution
using \texttt{CMFT} tracking.  Here increasing $t$ corresponds to
propagation
into the past. The $t$ used here increases into the future after the
merger.
} \label{g:0}
\end{figure}
Figure (\ref{g:0}) shows an order of magnitude in the evolution of
$\rho$ just after the topological transition, using the \texttt{cmft}
code.
For $0.2 \leq \left\|t -
t^{*}\right\| \leq 0.6$ the radius of the throat appears to vary
exponentially, while for $\left\|t - t^{*}\right\| \approx 0.0$
the radius appears to vary linearly.  Figure (\ref{g:3}) shows
that behavior of $\rho$  is indeed purely exponential in $t$ far from
the topological transition. Further, figure (\ref{g:4}) (also found
using the
\texttt{cmft} code) shows evidence
near the transition for power law (in fact, linear)
scaling of the throat radius. Note that without arbitrarily high
resolution
or nested adaptive mesh refinement any scaling that is present in
the solution cannot be resolved to machine epsilon. Therefore, the
results presented here using the \texttt{cmft} code can at best
establish qualitative evidence for such scaling. Figure
(\ref{g:4}) does indeed present such evidence that the solution
exhibits power law scaling in analogy to the bifurcation of fluid
droplets. Further, according to the results of the \texttt{cmft}
code, it is here conjectured that the Kastor - Traschen solution
scales like $\gamma \approx 1$. These results
indicate that the \texttt{cmft} method is capable of probing
into the fine structure of black hole event horizons undergoing a
topological transition. Using the results from the \texttt{cmft}
code, with $100^2$ points on the tracked surface, we obtain
the computational estimate $t^*=-1.454.$
\begin{figure} % Imported eps example.
\epsfxsize=8cm
\centerline{\epsfbox{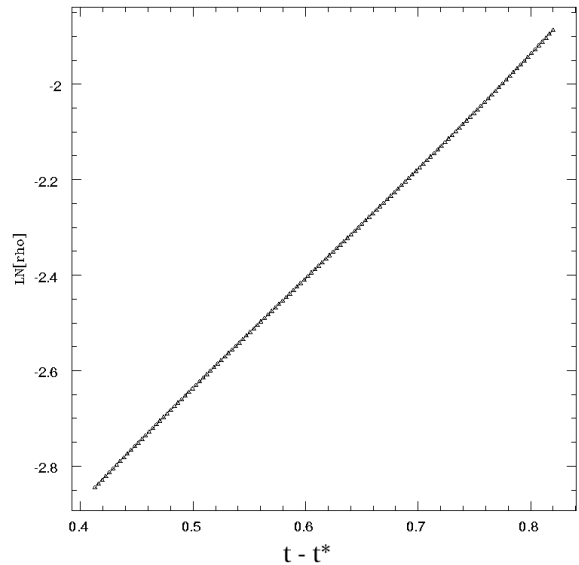}}
\vspace{0.5cm}
\caption{\texttt{CMFT} analysis of $e$ folding of the Kastor
Traschen solution during the part of the evolution after, and
away from the ``pinch on".
The $t$ used here increases into the future after the merger. }
\label{g:3}
\end{figure}

\begin{figure} % Imported eps example.
\epsfxsize=8cm
\centerline{\epsfbox{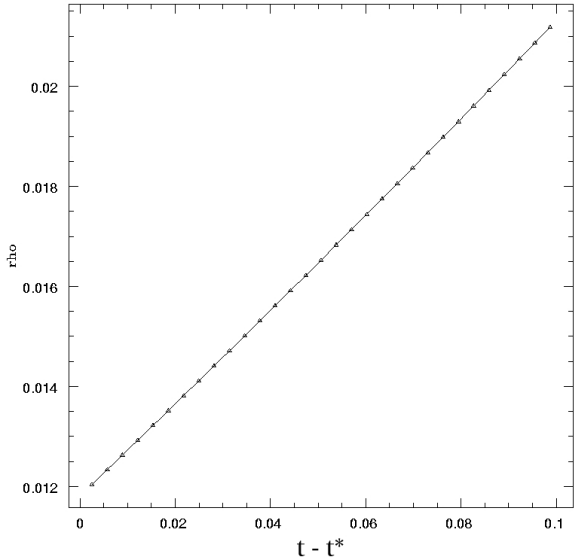}}
\vspace{0.5cm}
\caption{\texttt{CMFT} analysis of power law scaling at the
topological transition of the Kastor Traschen solution after the ``pinch
on".
The $t$ used here increases into the future after the merger.}
\label{g:4}
\end{figure}

\begin{figure}% Imported eps example.
\epsfxsize=8cm
\centerline{\epsfbox{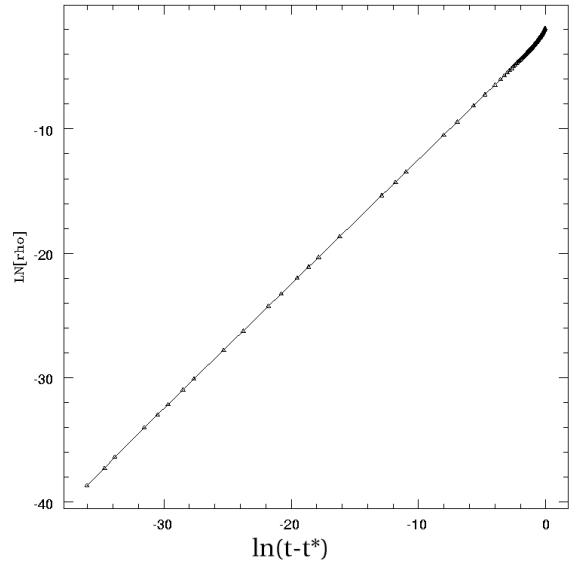}}
\vspace{0.5cm}
\caption{ Power
law scaling of topological transition in Kastor Traschen
solution after the ``pinch on", computed using an adaptive Runge Kutta
method.
The $t$ used here increases into the future after the merger.}
\label{i:1}
\end{figure}

This result is somewhat different from the analytical estimates of
Ref. \cite{kt2:}. However, we can verify this result because we can
carry out a much more accurate direct (ordinary differential equation)
integration to find the merge time. We use
the equatorial symmetry of this equal mass case merger, for which the
$\rho \approx t$ behavior is simply understood. In the equal mass
case the throat is circular in cross section, and this cross section
lies in the plane of symmetry of the problem. Suppose we introduce
a cylindrical coordinate system for the metric (\ref{kt}). The event
horizon
is defined by outgoing null rays, and the throat thus is generated by
null
rays moving in the $\rho$ direction (the throat moves at the speed of
light):
\begin{equation}
\frac{d\rho}{dt} =  U^{-2}
\end{equation}
where
\begin{equation}
U = H t + 2 M / \sqrt{\rho^{2} + a^{2}}.
\end{equation}
Figure (\ref{i:1}) shows a log vs log plot of $\rho$ vs $t-t^{*}$,
again begun from the late time single apparent horizon. It
indicates a critical exponent of exactly $\gamma = 1$.
This plot was generated using an adaptive, Runge Kutta , numerical
integration of the ordinary differential equation for
$\rho\left(t\right)$ to drive the difference $t^{*} - t$ to
below machine epsilon. The $t^*$ found in this simulation is
$t^* = 1.45845468374$, in agreement within accuracy to the \texttt{cmft}
result.

Straightforward physics explains this answer. The instant of merger
$t^{*}$ is not a special value for the Kastor - Traschen metric
function $U$. Hence, near this event, equation (35) for $\rho$ is
simply \begin{equation} \frac{d \rho}{dt} = U^{-2}\left(t^{*}\right) =
\mathrm{const} \neq 0 \end{equation} plus higher order terms. Thus, in
particular, the geometrically significant quantity, the circumference
of the throat, $C(t) \approx U(t^*) 2\pi\rho(t)$ grows linearly with
time, $\gamma = 1 $.

\section{Conclusions}
We have argued here that adaptive mesh refinement applied
to the front tracking approach to tracking black hole
event horizons is a potentially fruitful method;
and moreover proves necessary for black hole processes in which
the event horizon undergoes either a change of topology or develops
creases or caustics.
Towards that end we have developed an adaptive mesh technique
that makes use of comoving meshes and demonstrated the
accuracy of our approach. Applying our method to the symmetric
collision data of the Kastor Traschen analytic solutions, we
have found the surprising result of a power law in the minimal
radius of the throat connecting the black holes following merger.
We have shown power law scaling for the throat diameter  at merger
of black holes in a special situation: $\Lambda =3 (\neq 0)$, $M_1=M_2=0.1
$,
$Q = M$, and axisymmetic.
This effect is in analogy to that of fluid droplet problems.
Since
it is generically true that the black hole event horizon satisfies
dynamical evolution equations that are exactly the form of a two dimensional
fluid, we conjecture and give some qualitative arguments below concerning
the specializations in the Kastor Traschen to argue that the power
law scaling found in the
Kastor Traschen solution must be generic for the symmetric collision
of two black holes. Extending the analogy further, we also expect
a similar power law for the asymmetric problem, although generalized
in that case to accomodate the asymmetry of the problem as we now argue.

It is clear that $\Lambda \neq 0 $ is irrelevant to this phenomena. The
scaling
behavior applies to an arbitrarily thin neck that first connects
merging black holes. The curvature of the neck scales as
$1/(\rho \times \mathrm{separation}) \approx \left(\rho M\right)^{-1}$.
For small $\rho M$ this dominates
the fixed contribution.

The assumption of extremal black holes ($Q = M$) is a strong one.
In Newtonian terms, it implies no acceleration of the motion
between the holes (the electrical force is of equal magnitude and opposite
the gravitational force). A first intuition  may be that this has something
to do with the exponent $\gamma$ in the scaling of the throat radius
($\gamma = 1$ both numerically and analytically). However, the generic
axisymmetric merger of black holes will deal with the scaling in a very
brief
interval at the beginning of merger; the ``relative velocities'' of the
holes
will be essentially constant in any case. We thus predict that
extremality  ($Q = M$) is not essential to the $\gamma = 1$ power law
behavior
found here. For similar reasons we do not expect the spin of holes
(zero in the case of the Kastor - Traschen solution) to affect the results
found here. [The special (non-generic) case where non-extremal equal mass
hole data are set with the holes just at merger {\it will} be different,
because
we expect the holes to ``accelerate from zero'' as the throat grows.]

Still in the context of axisymmetry, we can investigate whether the
assumption of equal mass is
relevant to the value of $\gamma$. In the
limit of $M >>m$ we essentially consider a null topological tube (the
small hole $m$) merging with a null plane $M$. We are investigating
this configuration for merger. We again expect power law scaling; the
question is to determine the exponent. If in this case $\gamma \neq 1$,
then $\gamma$ will in general be a function of the ratio of the masses:
$\gamma = \gamma(m/M)$; $\gamma(1) = 1$.  We no longer have the
symmetry plane to simplify analysis; the throat cannot be followed by
following the individual $\rho$ direction null trajectories, but
Lindblom \cite{Lindblom} has given an argument, based on the smoothness of
the throat
after its formation, that gives $\gamma = 1$ for the growth of the
circumference in this case also.

Finally, we consider relaxing axisymmetry. In analysis and simulations
based on perturbation of late time horizons evolved backwards\cite{wini2:},
toroidal  behavior is sometimes seen in the non-axisymmetric case.
The horizons grow more then one ``point'' as they approach merger; two
points
that touch simultaneously in very quick succession will create an
evanescent toroidal horizon.  To date, there has been no study directly
relating parameters of the toroid
to parameters of the colliding  holes (we have seen only suggestions of
toroidal
behavior in any of our
computations ({\it c.f.} the ``offset points" of the event horizon in
\cite{gr-qc0303099}, Figure 11) , perhaps because of resolution
limitations). However, we would
expect each merging ``point'' to behave like a seperate merger and thus to
undergo
power law scaling, but not necessarily
with $\gamma = 1$. In fact, analysis by Winicour \cite{wini2:} in the
generic
case shows that the toroid closes superluminally; thus prevents
sending signals ``through the hole''. This requires that the ``points'' grow
superluminally, which requires $\gamma < 1$. The crucial difference from
the case explained by Lindblom, is that the toroidal structure is not
described by a smooth surface. In particular, there is an inner
``crease" where geodesics are joining the torus, allowing it to close
superluminally.

\section{Ackmowledgment}
We thank L. Lindblom for very insightful and helpful comments. This
work was supported by NSF rant PHY0102204. Computations were done at the
University of Texas Advanced Computing Center.

\end{document}